\documentclass[12pt]{article}

\usepackage{amsfonts,amssymb,color,cancel,mathtools,setspace,tabularx}
\usepackage[margin=0.75in,letterpaper]{geometry}
 \usepackage{sidecap}
  \usepackage{amsmath}

\usepackage{graphicx} 
\graphicspath{ {images/} }

\usepackage{hyperref}
\hypersetup{
 colorlinks=true, 
    linktoc=page,     
    linkcolor=blue,
}
\newcommand{\br}[1]{\left[#1\right]}

\def\e{\epsilon}
\def\T{\Theta}

\newcommand{\bea}{\begin{eqnarray}}
\newcommand{\eea}{\end{eqnarray}}
\newcommand{\non}{\nonumber}
\renewcommand{\k}{\kappa}

\makeatletter

\renewcommand\section{\@startsection {section}{1}{\z@}%
  {-3.5ex \@plus -1ex \@minus -.2ex}%
  {2.3ex \@plus.2ex}%
  {\normalfont\large\bfseries}}

\renewcommand\subsection{\@startsection{subsection}{2}{\z@}%
  {-3.25ex\@plus -1ex \@minus -.2ex}%
  {1.5ex \@plus .2ex}%
  {\normalfont\bfseries}}

\begin{document}


\title{{ \LARGE Generalized Near Horizon Extreme Binary Black Hole Geometry}}

\author{
{\large Jacob Ciafre}$^1$\footnote{jakeciafre@outlook.com} , \
{\large Shahar Hadar}$^{2,4}$\footnote{shaharhadar@g.harvard.edu}  , \
{\large Erin Rickenbach}$^1$\footnote{rickenbacherin@gmail.com} , \
{\large Maria J. Rodriguez}$^{1,2,3}$\footnote{maria.rodriguez@aei.mpg.de; maria.rodriguez@usu.edu.}  \\
\\
$^{1}${\small  Department of Physics, Utah State University,}\\ {\small 4415 Old Main Hill Road, UT 84322, USA}\\
\\ 
$^{2}${\small Max Planck for Gravitational Physics - Albert Einstein Institute,}\\ { \small Am Muhlenberg 1, Potsdam 14476, Germany} \\
\\
$^{3}${\small Instituto de F\'isica Te\'orica IFT-UAM/CSIC,}\\ { \small 
C/ Nicol\'as Cabrera 13-15, Universidad Aut\'onoma de Madrid, 28049 Madrid, Spain} \\
\\
$^{4}${\small   Center for the Fundamental Laws of Nature, Harvard University,} \\ \vspace{0.1cm} { \small Cambridge, MA 02138, USA }
 }

\maketitle

\abstract

We present a new vacuum solution of Einstein's equations describing the near horizon region of two neutral, extreme (zero-temperature), co-rotating, {\it non-identical} Kerr black holes. The metric is stationary, asymptotically near horizon extremal Kerr (NHEK), and contains a localized massless strut along the symmetry axis between the black holes. In the deep infrared, it flows to two separate throats which we call ``pierced-NHEK'' geometries: each throat is NHEK pierced by a conical singularity. We find that in spite of the presence of the strut for the pierced-NHEK geometries the isometry group $\mathrm{SL(2,R) \times U(1)}$ is restored. We find the physical parameters and entropy.

\newpage

\tableofcontents
\section{Introduction}
Rapidly rotating, (near-)extreme Kerr Black Holes (BHs) constitute a unique arena which offers both observational relevance and enhanced theoretical control. Several high-spin candidates (c.f. \cite{Gou:2011nq}-\cite{McClintock:2006xd}) have been observed, and such BHs could produce  characteristic signatures for various current and future experiments, including gravitational-wave detectors such as LIGO/Virgo, and optical observatories such as the recently triumphant \cite{Akiyama:2019cqa} Event Horizon Telescope. Theoretically, (near-)extreme BHs are especially tractable since they develop an emergent conformal symmetry. More precisely, they admit a non-degenerate near-horizon geometry, the so-called near-horizon extreme Kerr (NHEK) geometry \cite{Bardeen:1999px}. This geometry is interesting: every fixed polar angle slice of it can be thought of either as 2-dimensional anti de-Sitter space ($\mathrm{AdS_2}$) with a circle nontrivially fibered upon it or (equivalently) as a quotient of the so-called warped $\mathrm{AdS_3}$ spacetime. Consequently it enhances the isometry group of Kerr, $\mathrm{R \times U(1)}$ (corresponding to stationarity and axissymetry), to $\mathrm{SL(2,R) \times U(1)}$. This motivated the Kerr/CFT conjecture \cite{Guica:2008mu}, which hypothesizes that the Kerr BH is dual to a ($1+1$ dimensional) conformal field theory (CFT) living on the boundary of this near horizon geometry. This boundary can be thought of as the spacetime region in which the NHEK geometry is glued to the external, asymptotically flat, Kerr spacetime. 

The NHEK geometry has a simpler cousin---the Robinson-Bertotti universe or $\mathrm{AdS_2 \times S^2}$. This spacetime arises as an analogous near-horizon limit of maximally charged Reissner-Nordstr{\"o}m BHs. This type of BHs can be used to construct, remarkably simply, multi-BH configurations \cite{Hartle:1972ya}. Those are static solutions to Einstein-Maxwell theory with an arbitrary number of maximally charged (all with the same sign), non-rotating BHs of any mass. The time-independence of these solutions is possible since the BHs' gravitational attraction and electric repulsion cancel each other precisely---in the full nonlinear theory---for arbitrary BH positions. A neat observation regarding these solutions was made in \cite{Maldacena:1998uz}. Consider a system of two such maximally charged BHs. When they are widely separated, there exist also well-separated near-horizon (approximately $\mathrm{AdS_2 \times S^2}$) throats surrounding each one of the BHs. When the BHs are close to each other (relative to a length scale defined by a characteristic mass), however, there exists a region around them which is approximately an $\mathrm{AdS_2 \times S^2}$ throat which surrounds \emph{both} horizons, and only when moving further towards either one of the horizons does one recover the two separate throats. This phenomenon was coined in \cite{Maldacena:1998uz} ``AdS fragmentation'': the joint throat fragments into two smaller ones, when moving deeper into the infrared. This generalizes to an arbitrary number of throats: one ``trunk'' throat can fragment into several branches which can then   branch again, and so forth.

This compelling picture depends strongly on the properties of the special system of choice. The fact that it can be embedded in a supersymmetric theory, as a solution which preserves some supersymmetry \cite{Gibbons:1982fy}, guarantees this type of behavior. In this paper, we propose the closest possible analogue, presumably, to fragmentation in the case of maximally rotating, uncharged BHs. Since these are not supersymmetric anymore and there is no known smooth stationary solution involving such BHs, we allow for a conical singularity between the BHs which balances the gravitational attraction and keeps the system stationary. We study a 1-parameter family of exact axis-symetric solutions describing two corotating extreme Kerr BHs of arbitrary masses which are held apart by a conical singularity with effective pressure, usually called a \emph{strut} and as we rescale coordinates to zoom-in on the near-horizon region, we also shorten the strut separating the BHs. In this way we construct the exact solution corresponding to the region where NHEK fragments into two NHEK-like throats which are held apart by the strut. We call these ``NHEK2'' geometries. The solution presented here generalizes \cite{Ciafre:2018jpe}, which studied a similar construction for the equal-mass case. These infrared near-horizon geometries which the strut pierces on its way to the horizons are analogues of NHEK which include a conical singularity at one of the poles, extending from the horizon all the way to the NHEK boundary. We verify that this does not ruin the symmetry structure: the ``pierced-NHEK'' geometry still has an $\mathrm{SL(2,R) \times U(1)}$ isometry group. So while the full NHEK2 does not have $\mathrm{SL(2,R) \times U(1)}$, it interpolates from a geometry that does have conformal symmetry in the ultraviolet to two throats that are also conformally symmetric, in the infrared. 

Introducing conical singularities has caveats which are important to stress. First, the stability, both classical and quantum mechanical, of these solutions is questionable. A second point is that the type of conical singularities we use here, the struts, are of excess angle type (rather than deficit angle); the effective stress-energy associated to such objects has negative energy density. Keeping these caveats in mind, we still hope that this construction may be useful in various contexts. Firstly, such stationary BH binary solutions have been recently applied to study astrophysically motivated problems involving dynamical binaries (see for example \cite{Camps:2017gxz} for the use of quasi-stationary, extremally charged solutions in a gravitational-wave application); even though the physics governing the dynamics of these systems is different it was argued in \cite{Cunha:2018gql} (see also references therein) that in some cases such solutions can be used as tools for modeling the astrophysical systems' observational signatures, e.g. gravitational lensing. And secondly, these solutions may give some insight in the holographic, Kerr/CFT context. In this regard, it is interesting to note a recent study of holography and thermodynamics with conical singularities in the bulk \cite{Anabalon:2018ydc}. It should be possible to generalize our construction to an arbitrary number of BHs with arbitrary masses.

The workhorse of this paper are the binary BH solutions first found in \cite{Kinnersley:1978pz} and further studied, including their construction via various solution generating techniques in \cite{Kramer}-\cite{Manko:2011ts}. These exact solutions are stationary, axisymmetric, asymptotically flat solutions which describe two rotating BHs held apart by a strut along the symmetry axis. The BHs of these solutions can have arbitrary masses and spins and in particular can be either co- or counter-rotating. We are interested in the case in which the BHs are maximally co-rotating, with arbitrary masses. In particular, we start from the corotating solution described in \cite{Manko:2011qh}, and for the convenience of the interested reader we describe it explicitly  in the so-called Weyl-coordinates in Appendix \ref{appendix:full solution}. This coordinate choice serves best to describe classes of stationary and axisymmetric solutions of Einstein's Theory of General Relativity in vacuum. 

The rest of this paper is organized as follows.   We first construct the new Generalized Near Horizon Geometry of the stationary binary extreme-Kerr BH solution in section \ref{sec:GenNHEK2} and analyze its physical properties. In particular, we show how it admits a localized strut along the symmetry axis between the black holes but is asymptotically NHEK. In Section \ref{sec:NHEKwithcone} we zoom-in further to the infrared of each throat, and find the near-horizon geometries in which the strut pierces the horizons, extending from the horizon all the way to the NHEK boundary. We show that in spite of the strut, the pierced-NHEK geometries have an $\mathrm{SL(2,R) \times U(1)}$ isometry group. Finally, we will summarize the key results of the paper in Section \ref{sec:Discussion}.

\section{Generalized-NHEK2: Generalized Near Horizon Geometry of Extreme Binary Kerr Black Holes Solution}
\label{sec:GenNHEK2}

In this section, we construct the Generalized Near Horizon Geometry of Extreme Binary Kerr (Generalized-NHEK2) black hole solution. Our starting point, is the stationary solution to Einstein equations in vaccum \cite{Manko:2011qh} that contains two extremal (zero-temperature) co-rotating black holes. For convenience   and for fixing the notation, we reproduced the original results of \cite{Manko:2011qh}  in Appendix \ref{appendix:full solution}. We will only consider the solutions characterized by positive values of the mass that correspond to the parameter range
\begin{eqnarray}\label{parameterp}
-\frac{1}{\sqrt{2}}\le p<0, \qquad q>0,\qquad q<P\le 1\,.
\end{eqnarray}
%
Note that for $P=+1$ the equal mass case, treated in \cite{Manko:2011ts}, \cite{Ciafre:2018jpe}, is recovered \footnote{As described in \cite{Manko:2011qh}, there is another solution with positive mass that corresponds to $-1<P<-p$. This solution belongs to a more problematic case containing a massless ring singularity outside the symmetry axis that we will not consider here.}; the extreme mass ratio limit is recovered for $P \to (\sqrt{1-p^2})_+$ or $P \to (-p)_-$ . 

\subsection{Near-horizon limiting procedure}
\label{genNHEK2metric}

In previous works \cite{Ciafre:2018jpe} we developed the necessary tools to inspect the extreme co-rotating binary Kerr black hole solution. This section is nevertheless self contained. We proceed to compute the near horizon geometry of extremal nonidentical binary Kerr black holes solution, that we are going to refer to as  ``Generalized-NHEK2''. 

The solution of extremal BBHs \cite{Manko:2011qh} -  that we reproduced in Appendix \ref{appendix:full solution} - has a rather more compact representation in Weyl coordinates. We therefore perform the scaling computations in these coordinates. In this case, we find that the appropriate near-horizon limiting procedure for the extremal BBHs is
\begin{eqnarray}\label{rescalings1}
\rho =  \hat{\rho}\, \lambda  \,,\qquad z =  \hat{z}\,\lambda \ ,\qquad t=\frac{\hat{t}}{\lambda}\,,\qquad \phi=\hat{\phi}+\frac{1}{2 M}\frac{\hat{t}}{\lambda}\,,
\end{eqnarray}
\begin{eqnarray}\label{rescalings2}
p= -\frac{1}{\sqrt{2}}+\frac{3 \sqrt{2}-2 P}{4}\, \lambda \, , \qquad \kappa =  M \, \lambda\, .
\end{eqnarray}
%
Taking $\lambda \rightarrow 0$ and keeping $( \hat{t},  \hat{\rho},  \hat{z},  \hat{\phi})$ fixed.
As a result of this procedure, we find the generalized (nonidentical mass) Generalized-NHEK2 geometry 
\begin{equation}\label{NHEK2sol1}
ds^2=-\frac{\hat{\rho}^2}{f}d\hat{t}^2+f (d\hat{\phi}+\omega \,d\hat{t})^2+e^{2\nu}(d\hat{\rho}^2+d\hat{z}^2) \, ,
\end{equation}
defined by the equations
\begin{small}
\begin{eqnarray}
\label{NHEK2sol2}
f&=&-\frac{4 M^2\mu_0\,(\mu_0+ 2\sigma_0^2)}{ \mu_0\,(\mu_0+2\sigma_0^2-2\sigma_1+\pi_0)+\mu_1\, \pi_1+(1-y^2)\,\sigma_0\,\tau_0},\, \nonumber \\
\omega&=&-\frac{ \pi_0 \,\sigma_0+\pi_{1}\,\sigma_1-\mu_1-4 \sigma_0 \,\sigma_1-(1-y^2)\,\tau_0/2}{2M(\mu_0+2\, \sigma_0^2)}\,,\\
e^{2\nu}&=&\frac{ \mu_0\,(\mu_0+2\sigma_0^2-2\sigma_1+\pi_0)+\mu_1\, \pi_1+(1-y^2)\,\sigma_0\,\tau_0}{K_0^2\,(x^2-y^2)^4}\,, \nonumber
\end{eqnarray}
\end{small}
where
\bea\label{functionsNHEK2}
\mu_0 &=& -\frac{\hat{\rho}^2}{2 M^2}\,,\qquad \sigma_0= -\frac{x^2-y^2}{2}+\beta_0 (x^2+y^2)-2\alpha_0\,  x\, y\,,\\
\pi_1&=&4 x (\beta_0\, x-\alpha_0 \,y)-(1+2\beta_0)( x^2- y^2)\,,\\
\mu_1&=&-\Delta_0(-1+x^2)^2+\frac{\sqrt{2} Q \beta_0^2}{\alpha_0}(x^2-y^2)^2\,,\\
 {\sigma_1}&=&\Delta_0(x^2-y^2)+(-2 \Delta_0\beta_0+\beta_1)(x^2+y^2)+2(2 \Delta_0 \alpha_0 {-}\alpha_1) x y\,,\\
\pi_0&=&\left[ 1-4 \Delta_0 \beta_0+\beta_1+\frac{ \beta_0 (\beta_1-K_1)}{K_0} \right]4x^2-\sqrt{2} (P x(1+x^2)-Q y (1+y^2))\\
&+&\left(\sqrt{2} (\beta_0 P+\alpha_0 Q)x-\sqrt{2} (\alpha_0 P+\beta_0 Q) y -\beta_3\right)(x^2-y^2)\,\\
&+&\left[4 \Delta_0\alpha_0- \alpha_1-\frac{ \alpha_0 (\beta_1-K_1)}{K_0}\right] 4 x y\,,\\ 
 {\tau_0}&=& \sqrt{2} (P x+ Q y)(x^2-1)+\left(\alpha_3+\frac{Q}{\sqrt{2}\alpha_0} x\right)(x^2-y^2) {-}\left(\alpha_3+\frac{Q}{\sqrt{2}\alpha_0}\right)(1-y^2)\, ,
\eea
where we use prolate spheroidal coordinates
\bea
x&=&\frac{\sqrt{\hat{\rho}^2+(\hat{z}+M)^2}+\sqrt{\hat{\rho}^2+(\hat{z}-M)^2}}{2M}  \, ,\\
 y&=&\frac{\sqrt{\hat{\rho}^2+(\hat{z}+M)^2}-\sqrt{\hat{\rho}^2+(\hat{z}-M)^2}}{2M}  \, ,
\eea
and introduce the notation
\begin{gather}
Q=\sqrt{1-P^2}\,,\qquad      \Delta_0=\frac{3 -\sqrt{2} P}{2} \,,\qquad  K_0=\beta_0-1/2\,,\\
    K_1=-\frac{1}{4} \left(\Delta_0-\frac{4}{\Delta_0}+7\right) (1-2 \beta_0)-\frac{2 \beta_1 K_0}{1-2 \beta_0}-4 \Delta_0 K_0\,,\\
      \alpha_0=\frac{Q}{\sqrt{2}-2 P}\,,\qquad  {\beta_0}= {-}\sqrt{\alpha_0^2+1/2}\,\\
   \alpha_1=\frac{2 Q^2\beta_0^2}{\alpha_0}-\frac{\Delta_0}{\sqrt{2}}\,Q (1+4 \beta_0^2)\,,\qquad \beta_1=\frac{2 P Q\beta_0^2}{\alpha_0}-\frac{\Delta_0}{\sqrt{2}}(P+4 Q \alpha_0\beta_0)\,,\\
 {\alpha_3}= {-}\left(\frac{5 \Delta_0}{{2}}-\frac{{2}}{\Delta_0}+\frac{3}{2}\right)\,,\qquad
\beta_3=\frac{\Delta_0 \left(1-8 \beta_0^2\right)+2 (\alpha_0\alpha_1+ \beta_0 \beta_1)}{K_0}+\frac{\left(1-4 \beta _0^2 \right) K_1}{2 K_0^2}\, ,
\end{gather}
for $1/\sqrt{2}<P\le1$.

\subsection{Physical Parameters}

Let us now consider the physical parameters of the Generalized-NHEK2 solution. As we did at the level of the geometry, the near-horizon limiting procedure can be applied to the original physical parameters found in \cite{Manko:2011qh} (also reviewed here in App. \ref{appendix:full solution}). Applying this technique yields the expressions for the masses $M_{1}, M_{2}$, and angular momenta $J_{1},J_{2}$ in the Generalized-NHEK2 solution
\bea
M_{1} =\frac{M}{2}\left(1- \frac{Q}{\sqrt{2}-P}\right), \qquad M_{2 }=\frac{M}{2}\left(1+ \frac{Q}{\sqrt{2}-P}\right) \, ,
 \eea
\bea
J_{1 }=2 M_1^2 \left(1- \frac{Q}{\sqrt{2}-P}\right)^{-1} ,\qquad J_{2 }=2 M_2^2 \left(1+ \frac{Q}{\sqrt{2}-P}\right)^{-1},   
  \eea
and the corresponding angular velocities 
\bea
\Omega_1= \Omega_2=\frac{1}{2\,M}  \,.
\eea
satisfying at the same time the Smarr relation $M_1=2\,J_1 \Omega_1$ and $M_2=2\,J_2 \Omega_2$. It is worth noticing that the new solution contains objects that are in thermal equilibrium. The black hole entropy is, as usual, the area of the event horizon divided by 4. This gives
\bea
S_1=4 \pi  M^2 \left(2-\sqrt{2} (P+ Q)\right)\,,\qquad S_2=4 \pi  M^2 \left(2-\sqrt{2} (P-Q)\right)\,.
\eea
%
\subsection{Ergospheres}

The Generalized-NHEK2 spacetime that we constructed contains regions where the vector $\partial_t$ becomes
null. We will refer to the boundary region as the ergosphere, since they are inherited from the presence of such regions in the original stationary extreme BBHs geometries. For NHEK2 these are defined by regions where $g^{tt} = 0$ and give rise to a set of disconnected regions as shown in Fig. 1. Different values of the parameter $P$ are bounded by the extreme mass ratio solution when  $P=1/\sqrt{2}$ and identical mass solution when  $P=1$. The horizons of the black holes in Generalized-NHEK2 are points in the $(\hat{\rho}, \hat{z})$-plane and have finite horizon areas. There is a self similar behavior close to each black hole that resembles the ergospheres' diagrams of isolated extremal Kerr black holes  .

\begin{figure}[h!]
\begin{center}
\includegraphics[width=4.8cm,height=4.8cm]{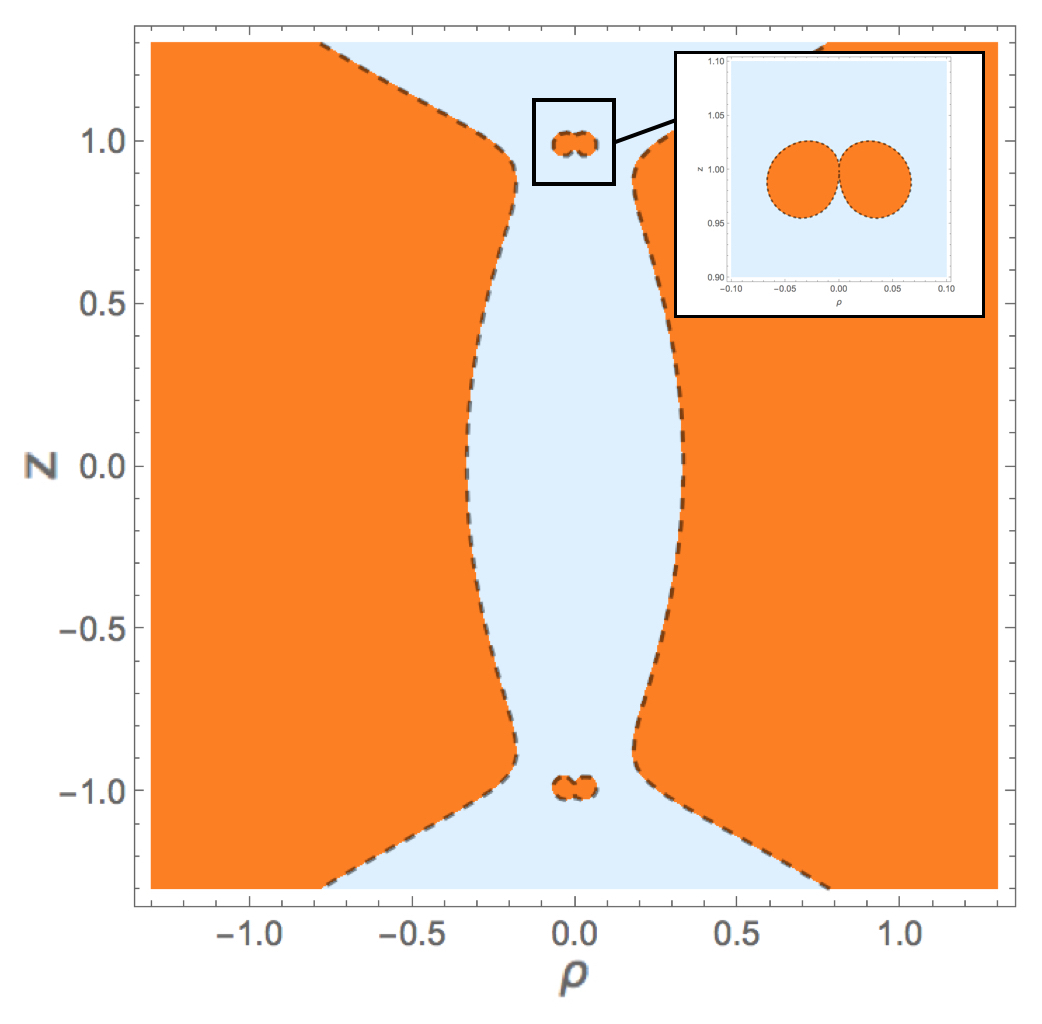}
\includegraphics[width=4.8cm,height=4.8cm]{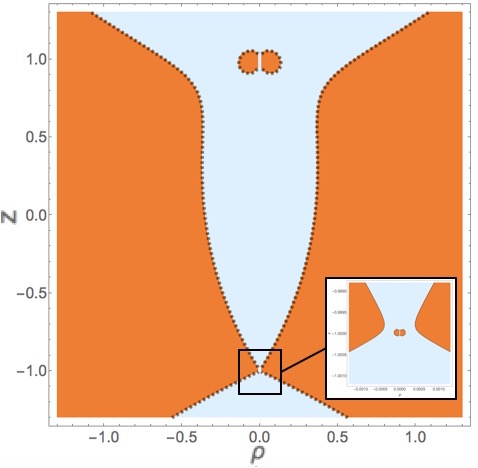}
\includegraphics[width=4.8cm,height=4.8cm]{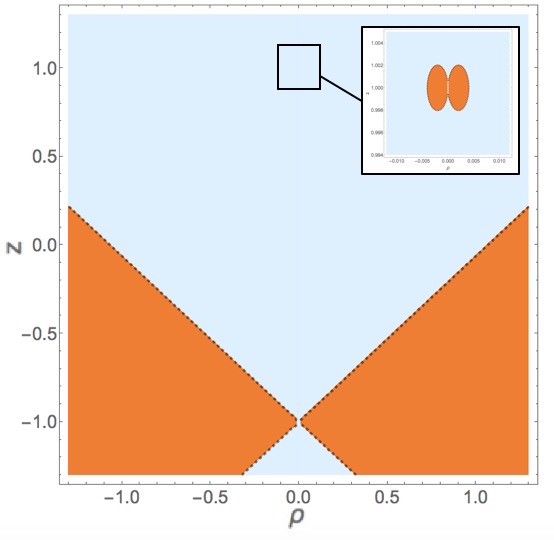}
\caption{\small Ergoregion (shaded {\it orange region}) of the Generalized-NHEK2-black hole solution for $M=1$ and $P=1,0.9,0.71$ (from {\it left} to {\it right}). Each black hole is located at $\rho=\hat{\rho}_H=0, z=\hat{z}_H=\pm 1$.   Magnified diagrams, close to the locations of the black holes appear in the corners  . The dashed line corresponds to the boundary where $\partial_{\hat{t}}$ is null. The different values of $P$ in the diagrams correspond to solutions with BBHs of distinctive mass ratios. Note that  in the Generalized-NHEK2 solution the parameter  $1/\sqrt{2}<P\le1$ where $P=1$ is the equal mass identical black hole case, and  $1/\sqrt{2}\sim 0.07071$ the extreme mass ratio limit.}
\label{ergo}
\end{center}
\end{figure}

\subsection{Asymptotic behavior}
In the asymptotic limit, for $\hat\rho=r\sin\theta$, $\hat{z}=r\cos\theta$ and $r\rightarrow\infty$, the Generalized-NHEK2 geometry in Section \ref{genNHEK2metric} has a limiting metric that corresponds to the NHEK metric - in Weyl coordinates - (\ref{NHEK2sol1}) with functions
\begin{gather}\label{Weylalt}
f=\frac{4 M^2 \hat{\rho} ^2}{2 \hat{z}^2+\hat{\rho} ^2}\, ,\qquad \omega=\frac{\sqrt{\hat{z}^2+\hat{\rho} ^2}}{2 M^2} \,,\quad
 e^{2\nu}=\frac{M^2 \left(2 \hat{z}^2+\hat{\rho} ^2\right) }{\left(\hat{z}^2+\hat{\rho} ^2\right)^2}\, .
\end{gather}
In other words, the Generalized-NHEK2 solution is asymptotically NHEK. 
It is worthwhile to mention at this point that in \cite{Amsel:2009ev,Dias:2009ex} it was shown that in the case of 4D Einstein gravity the NHEK geometry is the unique (up to diffeomorphisms) regular stationary and axisymmetric solution asymptotic to NHEK with a smooth horizon. The NHEK2 geometry that we unveil is asymptotically NHEK, but is not diffeomorphic to NHEK; this is not in contradiction with the results of \cite{Amsel:2009ev,Dias:2009ex} since the NHEK2 geometry is not smooth on the strut which keeps the BHs apart.

\subsection{Conical singularity}
As we have shown in   the previous subsection, the Generalized-NHEK2 is exactly asymptotically NHEK without any conical defects. However, as in the original stationary, extremal BBHs geometry there is in the bulk, a conical singularity on the $\hat{\rho}=0$ axis localized between the two black holes. In Weyl coordinates the conical singularities can be easily computed
\begin{equation}
\Delta\hat{\phi}= 2\pi \lim_{\hat{\rho}\rightarrow 0} \left(1-\sqrt{\frac{f}{\hat{\rho}^2 e^{2\nu}} }\right)\,, \quad -M<z<M \,,
\end{equation}
Our computation for the Generalized-NHEK2 metric shows that there is a non-removable conical excess between the two horizons.
\begin{equation}\label{conical}
\Delta\hat{\phi} =2\pi\,\left(1-\frac{1}{2(\sqrt{2}-P)^2} \right)\,.
\end{equation}
Outside this localized conical singularity our solutions are smooth.  

\section{Pierced-NHEK: near horizon limit at finite separation}
\label{sec:NHEKwithcone}

In this section it is shown that there exists a well-defined near-horizon limit of the stationary binary extreme Kerr solution \cite{Manko:2011qh,Manko:2011ts} even when the BHs, which are held apart by a conical singularity, are separated by a finite distance. The near-horizon region is composed of two disconnected NHEK-like geometries, one near each of the BHs. Each such geometry can be thought of as ``NHEK pierced by a cosmic string", the strength of which is determined by the distance between the BHs. The cosmic string/conical singularity balances the gravitational attraction of the companion BH, thereby enabling stationarity. The cosmic string extends all the way from the horizon to infinity in this geometry which we call the ``pierced-NHEK".  

Our starting point is the solution given in \cite{Manko:2011ts} (that corresponds to the identical mass binary black hole metric in \cite{Manko:2011qh} for $P=1$ which for convenience we reviewed in Appendix \ref{appendix:full solution}). As the most general solution is quite involved, we will start by fixing the parameters at a specific, convenient value which will be enough to convey our point regarding the existence of a nonsingular near-horizon geometry. It could be nice to explicitly write down the full most general expression, for arbitrary value of $P$, but for the sake of simplicity we will only focus on the $P=1$ case.

Starting with the solution presented in \cite{Manko:2011ts} with parameters $p=-1/2$, $\kappa = 4(\sqrt{33}-1)^{-1}$ (which sets\textsl{•} $M=1$, $J=2$) and coordinates denoted by $\{{\rho},{z},{t},{\phi}\}$, we choose to focus on the BH located at $\tilde{z}=\kappa$ and use the transformation
\begin{eqnarray}
{\rho} =  \e  \, R \sin \T  \,,\qquad {z}-\kappa =  \e \, R \cos \T \ ,\\
  {t} = \frac{17+\sqrt{33}}{16} \, \, \frac{T}{\e}\,,\qquad {\phi}= \left(\Phi + \omega_0 \, \, \frac{17+\sqrt{33}}{16} \, \frac{T}{\e}\right)\,,
\end{eqnarray}
where  $\omega_0 = (\sqrt{11}-\sqrt{3})/4$
facilitates the transition into a frame which co-rotates with the BH. Taking $\epsilon \to 0$ yields the nonsingular geometry
\begin{eqnarray}
ds^2=\Gamma(\T)\br{-R^2\,d T^2+\frac{d R^2}{R^2}+d\T^2+\Lambda^2(\T)\left(d\Phi+ \frac{\sqrt{11}-\sqrt{3}}{2} R\,d T\right)^2},
\label{conical NHEK metric}
\end{eqnarray}
where
\begin{eqnarray}
	\Gamma(\T) &=& \frac{2\left(3\sqrt{33}-13\right)\cos\T+\left(15-\sqrt{33}\right)(3+\cos2\T)}{16},\nonumber \\ \nonumber \\
	\Gamma(\T) \, \Lambda(\T)^2 &=&  \frac{256  \sin^2\T}{  4 \left(-59 + 11 \sqrt{33}\right) \cos\T + \left(93 - 
     13 \sqrt{33}\right) (3 + \cos2\T)} \, .
\end{eqnarray}

In this geometry, a priori, there could be a conical singularity either at $\T=0$ or at $\T=\pi$. Using (\ref{conical}), however, shows explicitly that at $\T=0$ there is no conical singularity whereas for $\theta=\pi$ there is an angular excess of 
\begin{equation}
{\Delta\Phi} =2\pi \, \frac{\sqrt{33}-1}{8} \,.
\end{equation}

Writing the pierced-NHEK geometry in the form (\ref{conical NHEK metric}) shows immidiately that it enjoys the isometry group $\mathrm{SL(2,R) \times U(1)}$, just like NHEK: the strut on the symmetry axis does not spoil this symmetry.

\section{Discussion}
\label{sec:Discussion}

\begin{figure}\label{figure}
 \begin{center}
\includegraphics[width=0.6\textwidth]{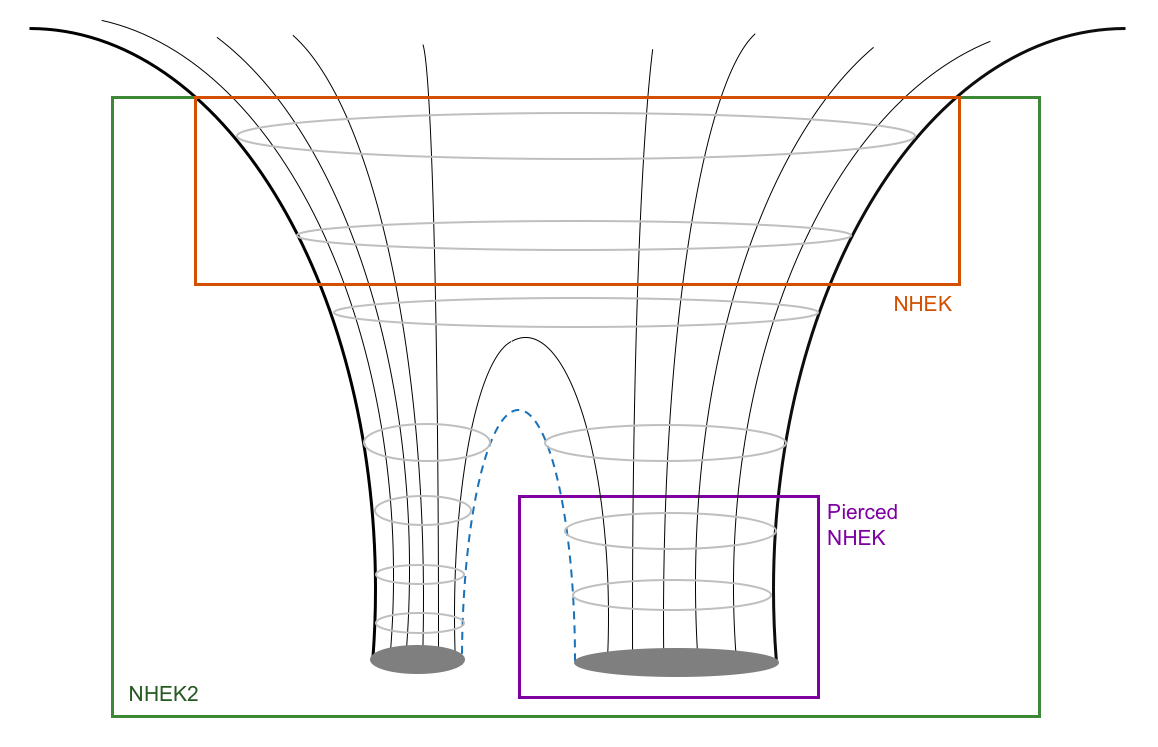}
\caption{The diagram represents a spatial cross section of the two extremal co-rotating BHs metric \cite{Manko:2011qh} reproduced in Appendix \ref{appendix:full solution}. The geometry (in {\it black} and {\it gray}) has an asymptotically Minkowskian region and a single black hole throat of mass $M_1+M_2$ which divides into two throats of masses $M_1$ and $M_2$. The   strut (conical singularity) in the solution is localized between the two black holes ({\it dashed blue}). In the infrared limit $\lambda \rightarrow 0$, when zooming into the near horizon limit, the throat becomes infinitely long and the Minkowski region decouples. This is the new Generalized NHEK2 solution that we constructed (represented in {\it green}), which is asymptotically NHEK (in {\it orange}). The splitting of the throat into two pieces survives this limit. In the deep infrared when zooming close to one of the horizons, we find new geometries that we call 'pierced NHEK'. These correspond to the NHEK metric pierced by a conical singularity on the symmetry axis, which runs from one of the poles up to the boundary (in {\it purple}).}
 \end{center}
\end{figure}

The aim of this paper was to unveil and analyze the Generalized-NHEK2 geometry. This geometry is obtained via a limiting procedure that we developed: a zoom-in on the near-horizon region of a 1-parameter family of corotating\footnote{The counter-rotating counterpart cannot be used to construct a similar solution since asymptotically it appears as a non-extreme BH.}, double-extreme Kerr solutions of arbitrary masses where the two BHs are parametrically close to each other and are held apart by a conical singularity (strut). The distance between the BHs is scaled to zero at the same rate of the zoom-in on the near-horizon region. This gives a relatively simple solution, which is asymptotically NHEK, and in the infrared flows to two separate throats which we call ``pierced-NHEK'' geometries: each of them is, approximately (when zooming further towards one of the horizons), NHEK pierced by a conical singularity on the symmetry axis, which runs from one of the poles up to the boundary. We find that in the deep infrared where the geometry is approximately ``pierced-NHEK'', the presence of the strut does not break the isometry group $\mathrm{SL(2,R) \times U(1)}$---it is restored there. In figure 2, we illustrate the structure of the Generalized-NHEK2 geometry.
The Generalized-NHEK2 solution asymptotes to NHEK, yet it is not diffeomorphic to NHEK. This is not in contradiction to the discussions in \cite{Amsel:2009et},\cite{Kunduri:2013ana} since in these papers, smoothness is assumed while here we allow for a conical singularity which balances the gravitational attraction between the BHs. This paper generalizes the construction studied recently in \cite{Ciafre:2018jpe} for the equal mass case.

\section*{Acknowledgements}

We would like to thank David Chow and Oscar Varela for helpful discussions. This work was supported by the NSF grant PHY-1707571 at Utah State University and, partially, by the Max Planck Gesellschaft through the Gravitation and Black Hole Theory Independent Research Group. SH gratefully acknowledges support by the Jacob Goldfield Postdoctoral Support Fund.  

\appendix


\section{Stationary extreme Kerr binary with strut - full solution}
\label{appendix:full solution}

Here we record, for completeness, the full exact solution corresponding to two extremal co-rotating BHs which are held apart by a strut lying on the joint rotation axis, between the BHs. The solution is axisymmetric, stationary and asymptotically flat. We follow the conventions of \cite{Manko:2011qh} in which this solution was presented.

Define prolate spheroidal coordinates $(x,y)$ by
\bea
x&=&\frac{\sqrt{\rho^2+(z+\kappa)^2}+\sqrt{\rho^2+(z-\kappa)^2}}{2\kappa}  \, ,\\
 y&=&\frac{\sqrt{\rho^2+(z+\kappa)^2}-\sqrt{\rho^2+(z-\kappa)^2}}{2\kappa}  \, ,
\eea
the metric is given by
\bea
ds^2=-\frac{{\rho}^2}{f}\,d{t}^2+f\, (d{\phi}+\omega\, d{t})^2+e^{2\nu}(d{\rho}^2+d{z}^2) \, ,
\label{manko and ruiz solution}
\eea
where:
\bea
f &=&\frac{\k(y^2-1)F}{D\,\omega}, \quad e^{2\nu}=\frac{D}{K_0^2(x^2-y^2)^4}, \quad \omega=-\frac{\k(y^2-1)F N}{[(\k(y^2-1)F)^2-\rho^2 D^2]}\, , \non \\
N &=& \mu^2-(x^2-1)(1-y^2) \sigma^2 \, , \non \\
D &=& N+\mu \pi+(1-y^2) \sigma \tau \, , \non \\
F &=& (x^2-1)\sigma \pi+\mu \tau \, ,\non \\
\mu &=& p^2(p^2 (x^2-1)^2 + q^2(1-y^2)^2 + (\alpha^2-\beta^2)(x^2-y^2)^2) \, , \non \\
\sigma &=& p^2(2\left[ p q(x^2-y^2) +\beta(x^2+y^2) -2\alpha x y \right]) \, , \non \\
\pi &=& p^2((4p^2/K_0) \left\{ (K_0/p^2)\left[ p P s
x(x^2+1)+2x^2+q Q y (y^2+1) \right]  \right. \non \\
 &+& \left. 2(p Q +p P \alpha+q Q \beta)\left[ p q y(x^2-y^2) +\beta y (x^2+y^2) -2\alpha x y^2 \right] \right. \non \\
 &-& \left. (K_0/p^2) (x^2-y^2) \left[ (p Q \alpha-q P \beta)x + (q P \alpha-p Q \beta)y \right] - 2(q^2 \alpha^2+p^2 \beta^2)(x^2-y^2) \right. \non \\
 &+& \left. 4(pq+\beta)(\beta x^2 - \alpha x y) \right\}) \, , \non \\
\tau &=& p^2( (4p^2/K_0) \left\{ (K_0/p^2) x \left[ (q Q \alpha + p P \beta)(x^2-y^2) - q P (1-y^2) \right]  \right. \non \\
 &+& \left.  ( p Q +p P \alpha + q Q \beta ) y \left[ (p^2-\alpha^2+\beta^2)(x^2-y^2)+y^2-1 \right]  \right. \non \\
 &-& \left. p Q (K_0/p^2) y (x^2-1) -2p(q \alpha^2 - q \beta^2 - p \beta)(x^2-y^2) -(pq+\beta)(1-y^2) \right\})\, , \non \\
K_0 &=& p^2(p^2+\alpha^2-\beta^2) \, ,
\eea
and the parameters are constrained so that
\bea
p^2+q^2=1 \, \, \, \, , \, \, \, \, P^2+Q^2=1 \, .
\eea 
For the corotating solution in which we are interested in this paper,
\bea
\alpha &=& -\frac{Q\left[ q \Delta+p q^2 + P(1+p^2) \right]}{2(p^2-Q^2)} \, , \non \\
\beta &=& \frac{p\left[ P \Delta + q(1+p P + Q^2) \right]}{2(p^2-Q^2)} \, , \\
\Delta &=& \sqrt{ 4p^2(1+pP)+q^2 (p+P)^2 } \, . \non
\eea

\subsection*{Physical Parameters}
The asymptotic metric does not contain a conical singularity, then the mass $M_{1},M_{2}$ and the angular momenta $J_1,J_2$ of the black holes can be easily calculated
\bea
M_1 &=& \frac{\kappa [ (q + p q P -p^2 Q) \Delta - (1 + p P)(p + p^3 + q^2 P -p q Q) + p q^3 Q] }{2 p (1+p P)(p^2-q^2)}, \non \\
M_2 &=&\frac{\kappa [ (q + p q P+p^2 Q) \Delta - (1 + p P)(p + p^3 + q^2 P+p q Q) -  p q^3 Q] }{2 p (1+p P)(p^2-q^2)}, \non \\
J_1 &=& \frac{(1 + p P + q Q ) M_1^2}{ 2(p + P)^2}[(1 + p P + q^2)\Delta-4 p q + p q (p-P)^2], \non \\
J_2 &=& \frac{(1 + p P - q Q ) M_2^2}{ 2(p + P)^2}[(1 + p P + q^2)\Delta-4 p q + p q (p-P)^2] \, ,
\eea
and, employing the Smarr relation, we can easily find the expressions for the angular velocities
\bea
\Omega_1=\frac{M_1}{2\,J_1} \,,\qquad \Omega_2=\frac{M_2}{2\,J_2} \,.
\eea

Additionally, the entropy for each black hole can be calculated to give:
\bea
S_1={\tfrac{2\pi p^2\kappa^2}{K_0^2}}\,(K_0(1+pP+qQ)-2p^2(\alpha-\beta)(p(Q+P\alpha+q)+(qQ+1)\beta))\,,
\eea
\bea
S_2={\tfrac{2\pi p^2\kappa^2}{K_0^2}}\,(K_0(1+pP-qQ)-2p^2(\alpha+\beta)(p(Q+P\alpha-q)+(qQ-1)\beta))\,.
\eea

\end{document}